%% file: main.tex
\documentclass[conference]{IEEEtran}
\IEEEoverridecommandlockouts

\usepackage{cite}
\usepackage{amsmath,amssymb,amsfonts}
\usepackage{algorithm}
\usepackage{algpseudocode}
\usepackage{graphicx}
\usepackage{textcomp}
\usepackage{xcolor}
\usepackage{booktabs}
\usepackage{subcaption}
\usepackage{bm}
\usepackage[hyphens]{url}
\usepackage{tikz}
\usepackage{stfloats}
\fnbelowfloat 
\usepackage[hidelinks]{hyperref}

\DeclareMathOperator{\tr}{tr}

\newcommand{\chol}{\text{chol}}

\definecolor{fusedcolor}{HTML}{E8899A}
\definecolor{rqblue}{RGB}{220,235,252}
\newcommand{\researchquestion}[1]{%
    \smallskip\noindent\fcolorbox{black}{rqblue}{\parbox{\dimexpr\linewidth-2\fboxsep-2\fboxrule}{\emph{#1}}}\smallskip}

\makeatletter
\def\bstctlcite{\@ifnextchar[{\@bstctlcite}{\@bstctlcite[@auxout]}}
\def\@bstctlcite[#1]#2{\@bsphack
  \@for\@citeb:=#2\do{%
    \edef\@citeb{\expandafter\@firstofone\@citeb}%
    \if@filesw\immediate\write\csname #1\endcsname{\string\citation{\@citeb}}\fi}%
  \@esphack}
\makeatother

\begin{document}
\bstctlcite{BSTcontrol}

\title{ADELIA: Automatic Differentiation for Efficient Laplace Inference Approximations}
\author{%
\IEEEauthorblockN{%
Afif Boudaoud\IEEEauthorrefmark{1},
Lisa Gaedke-Merzh\"auser\IEEEauthorrefmark{2},
Alexandros Nikolaos Ziogas\IEEEauthorrefmark{1},
Vincent Maillou\IEEEauthorrefmark{1},
Alexandru Calotoiu\IEEEauthorrefmark{1},\\
Marcin Copik\IEEEauthorrefmark{1},
H\r{a}vard Rue\IEEEauthorrefmark{2},
Mathieu Luisier\IEEEauthorrefmark{1},
Torsten Hoefler\IEEEauthorrefmark{1}%
}
\IEEEauthorblockA{%
\IEEEauthorrefmark{1}\textit{ETH Zurich}, Zurich, Switzerland \quad
\IEEEauthorrefmark{2}\textit{King Abdullah University of Science \& Technology (KAUST)}, Thuwal, Saudi Arabia%
}
\thanks{Corresponding author: Afif Boudaoud (afif.boudaoud@inf.ethz.ch).}%
}

\IEEEaftertitletext{\vspace{-10mm}%
\begin{center}
\refstepcounter{figure}\label{fig:overview}%
\includegraphics[width=\textwidth]{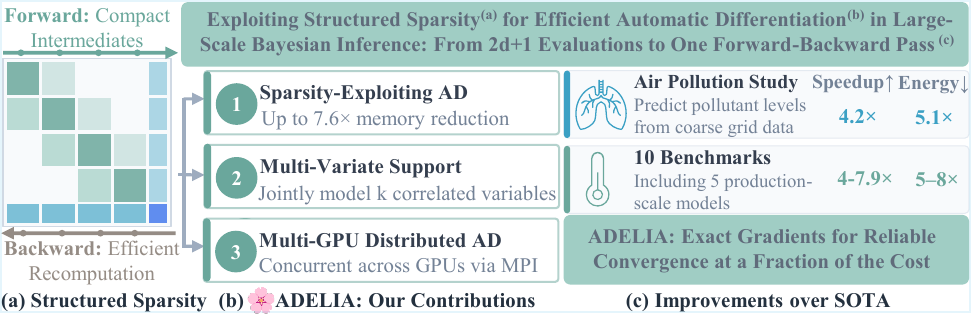}\\[1mm]
{\small Fig.~\thefigure. ADELIA replaces finite-difference gradients in DALIA~\cite{dalia_sc} with structure-exploiting reverse-mode AD.
(a)~Exploited sparsity pattern.
(b)~Key contributions enabling AD at HPC scale.
(c)~Per-gradient speedup and energy savings on real-world scale models.}
\end{center}
\vspace{-2mm}}

\maketitle

\begin{abstract}
\input{sections/abstract}
\end{abstract}

\begin{IEEEkeywords}
Automatic Differentiation, INLA 
\end{IEEEkeywords}

\input{sections/introduction}
\input{sections/background}
\input{sections/method}

\input{sections/evaluation}
\input{sections/related}
\input{sections/conclusion}

\section*{Acknowledgments}
All sections of this paper were iteratively refined using Claude by Anthropic for grammar and phrasing improvements.

\bibliographystyle{IEEEtran}
\bibliography{references}

\end{document}

%% file: sections/abstract.tex
Spatio-temporal Bayesian inference drives environmental and health sciences using latent Gaussian models. Integrated Nested Laplace Approximations (INLA) enable inference for these models at HPC scale but rely on derivative-based optimization over $d$ hyperparameters. State-of-the-art INLA implementations approximate derivatives via central finite differences (FD), requiring $2d{+}1$ evaluations. These evaluations are embarrassingly parallel, but total work and energy grow with $d$, limiting time-to-solution under fixed budgets. Reverse-mode automatic differentiation (AD) computes exact gradients independently of $d$, but its efficient application to INLA's structured-sparse kernels is an open challenge. We present ADELIA, the first AD-enabled INLA implementation with a structure-exploiting multi-GPU backward pass leveraging model sparsity. We evaluate ADELIA on ten benchmark models, including real-world air-pollution monitoring. We achieve $4.2$--$7.9\times$ per-gradient speedups and reliable convergence on production-scale models with up to 1.9M latent variables, where FD struggles. Even when scaled to 16--32 GPUs to match ADELIA's wall-clock time, FD consumes $5$--$8\times$ more energy.

%% file: sections/introduction.tex
\section{Introduction}
\label{sec:introduction}

Bayesian spatio-temporal models are widely used in climate, environmental, and biomedical sciences~\cite{gelman2013bayesian, wikle2019spatio, blangiardo2015spatial, moraga2019geospatial} to characterize complex processes and provide principled uncertainty estimates.
However, performing inference remains computationally challenging as the spatio-temporal structure typically gives rise to latent parameter space dimensions beyond the feasibility of traditional sampling-based approaches~\cite{gelman2013bayesian, brooks2011handbook}.
Integrated Nested Laplace Approximations (INLA)~\cite{rue2009approximate} addresses this issue by leveraging the conditional independence structure of the underlying latent Gaussian models, inducing block-tridiagonal arrowhead sparsity (BTA) patterns (Figure~\ref{fig:two_phase}a) in the associated spatio-temporal precision matrices~\cite{lindgren2022diffusion, gaedkeIntegrated2024}.
INLA’s computational cost is dominated by derivative computations in its optimization phase and during posterior approximation, which requires derivatives with respect to the model's hyperparameters.
State-of-the-art implementations, such as R-INLA~\cite{rue2009approximate} and DALIA~\cite{dalia_sc}, rely on central finite differences (FD), which require $2d+1$ function evaluations per gradient for $d$ hyperparameters and $2d^2+1$ evaluations for the Hessian during the post-optimization stage, incurring high computational costs and an inherent approximation error.
We consider both univariate and multivariate models, where multiple quantities are modeled jointly, raising the number of hyperparameters to $d=15$, further exacerbating the computational burden.

For scalar-valued functions, reverse-mode automatic differentiation (AD) computes exact gradients at $3$--$5\times$ the runtime cost of a single forward evaluation, regardless of~$d$~\cite{griewank2008evaluating, baydin2018automatic}.
Previous AD-based approaches to the Laplace approximations~\cite{kristensen2016tmb, margossian2020adjoint, margossian2023general} use general-purpose sparse solvers on CPU, leaving BTA structure unexploited at HPC scale.
For matrices with BTA sparsity, the forward pass preserves this sparsity, but the standard Cholesky derivative is dense, rendering it impractical. Attempting to preserve sparsity with standard AD techniques requires storing intermediate states across loop iterations, exceeding GPU memory at a million-variable scale.
A second challenge is posed by multivariate models, which introduce coupling terms between the $k$ response variables. Here, each block is a weighted combination of $k$ per-variable matrices, and the backward pass must decompose these to compute gradients for both per-variable and cross-variable hyperparameters.
Third, the largest models exceed single-GPU capacity, requiring the distribution of both the factorization and backward pass across multiple GPUs.

We present ADELIA\footnote{Code: \url{https://github.com/affifboudaoud/adelia-artifact}}, the first AD-enabled INLA framework for HPC-scale spatio-temporal inference (Figure~\ref{fig:overview}), and evaluate it on ten benchmark models, including real-world applications such as air pollution~\cite{dalia_sc} and temperature~\cite{gaedkeIntegrated2024} modeling. On the four multivariate models we test, FD not only slows convergence but prevents it, stalling at gradient norms orders of magnitude above tolerance; ADELIA's exact gradients resolve this, achieving well-conditioned optima while providing $4.2$--$7.9\times$ per-gradient speedups and $5$--$8\times$ energy savings on equal hardware.
These speedups combine algorithmic gains ($2d{+}1$ evaluations reduced to one) with framework-level efficiency differences, decomposed in Section~\ref{sec:eval:framework}.
Our main contributions are:

\begin{enumerate}
    \item \textbf{Structure-exploiting gradient computation for INLA.}
    We develop a custom differentiation rule for the full objective function, covering factorization, solve, and selected inversion of BTA precision matrices, that computes exact gradients while exploiting block sparsity. By fusing stages of the forward computation, only compact intermediate matrices need to be stored, reducing memory by approximately $2\times$.

    \item \textbf{Extension to multivariate models.}
    We extend the gradient computation to jointly model $k$ correlated variables by exploiting the block structure of the multivariate precision matrix, separating per-variable and cross-variable hyperparameter contributions to reduce per-step work (benchmarked up to $k{=}3$, $d{=}15$).

    \item \textbf{Multi-GPU distributed AD.}
    For models whose stored intermediates exceed single-GPU memory, we distribute the backward pass across multiple GPUs via temporal-domain decomposition with nested dissection reordering. We synchronize boundary intermediates via MPI allgather and stage intermediates on CPU memory, enabling AD on models up to 1.9M latent variables that otherwise cannot run on a single GPU.

    \item \textbf{Comprehensive evaluation on up to 128 GH200 GPUs.}
    Beyond reporting performance results, we provide a practitioner-oriented analysis that disentangles algorithmic gains from framework effects, evaluates when FD can match AD via scaling, and characterizes where time and memory concentrate in the gradient pipeline.
\end{enumerate}

%% file: sections/background.tex
\section{Background}
\label{sec:background}

\subsection{The INLA Pipeline}
\label{sec:bg:inla}

Environmental monitoring and public health policy rely on spatial and temporal predictions with quantified uncertainty: an air quality agency must know not only that pollution is elevated, but with what probability it exceeds a regulatory threshold; a climate service must attach confidence intervals to regional projections to inform adaptation planning.
Latent Gaussian models~\cite{rue2005gaussian} provide a principled Bayesian framework for these problems.
They consist of two layers of unknowns. The high-dimensional \emph{latent field}~$\bm{x}\in\mathbb{R}^N$ represents the unobserved spatio-temporal process of interest (e.g., pollution concentration across a fine mesh) and is modeled as a Gaussian Markov random field (GMRF) with sparse precision (inverse covariance) matrix~$\bm{Q}_p$. The second layer of unknowns is a small set of \emph{hyperparameters}~$\bm{\theta}\in\mathbb{R}^d$ ($d \ll N$) that governs global properties (spatial range, temporal range, observation precision).
Integrated nested Laplace approximations
(INLA)~\cite{rue2009approximate} make inference in these models
computationally feasible by replacing the expensive sampling of traditional Bayesian
methods (e.g., Markov chain Monte Carlo~\cite{gelman2013bayesian, brooks2011handbook}) with deterministic approximations that exploit the conditional independence structure of the GMRF.
This work targets all stages of INLA's algorithmic pipeline that require derivatives. They comprise two out of the three main stages and account for the majority of the computational workload, namely (1)~optimization of the scalar objective function~$f(\bm{\theta})$ over the hyperparameters~$\bm{\theta}$ using a quasi-Newton method, which requires repeated evaluations and gradients of~$f$; and
(2)~computation of the posterior Hessian at the optimum $\bm{\theta^*}$ to quantify hyperparameter uncertainty (Section~\ref{sec:method:hessian}).
Both stages rely on the same core linear algebra operations within ~$f$, which is defined as:
\begin{multline}
    f(\bm{\theta}) := \tfrac{1}{2}\log|\bm{Q}_p| - \tfrac{1}{2}\log|\bm{Q}_c| - \tfrac{1}{2}\bm{x}^{*T}\bm{Q}_p\bm{x}^* \\
    + \log \ell(\bm{y}\mid\bm{x}^*,\bm{\theta}) + \log p(\bm{\theta}) + \text{const}
    \label{eq:objective}
\end{multline}
Here $\bm{Q}_p(\bm{\theta})$ is the prior precision matrix and $\bm{Q}_c(\bm{\theta}) = \bm{Q}_p(\bm{\theta}) + \tau \bm{A}^T\bm{A}$ is the precision matrix after conditioning on observations under a Gaussian observation model, where $\tau$ is the observation precision, $\bm{y} \in \mathbb{R}^{n_y}$ denotes the observations, and $\bm{A} \in \mathbb{R}^{n_y \times N}$ is a sparse projection matrix mapping latent variables to observation locations.
The vector $\bm{x}^* = \bm{Q}_c^{-1}\bm{r}$, where $\bm{r} = \tau \bm{A}^T \bm{y}$, is the posterior mode (the solution of a linear system). 
The likelihood for the observations is denoted by $\ell(\bm{y}\mid\bm{x}^*,\bm{\theta})$ and $p(\bm{\theta})$ is the prior for the hyperparameters.

\textbf{Computational Structure:}
Evaluating $f(\bm{\theta})$ requires constructing $\bm{Q}_p(\bm{\theta})$ from sparse Kronecker products of spatial and temporal base matrices~\cite{lindgren2022diffusion, gaedkeIntegrated2024} and forming $\bm{Q}_c$;
Cholesky factorizations of both precision matrices for their log-determinants; a triangular solve for~$\bm{x}^*$; and a quadratic form $\bm{x}^{*T}\bm{Q}_p\bm{x}^*$.
The remaining terms ($\log \ell$, $\log p$) are cheap to evaluate once $\bm{x}^*$ is known.

\subsection{BTA structured Sparsity in Spatio-Temporal Models}
\label{sec:bg:structure}

\begin{figure*}[t]
    \centering
    \includegraphics[width=\textwidth]{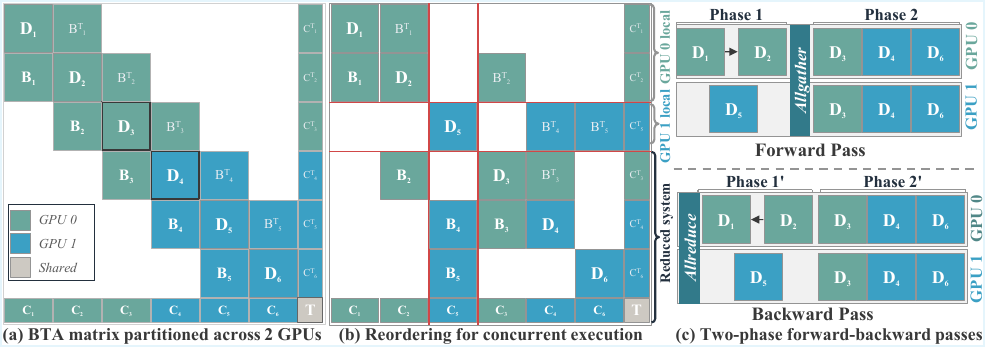}
    \vspace{-4mm}
    \caption{Distributed two-phase algorithm for $P{=}2$ GPUs on a BTA matrix with $n{=}6$ blocks. (a)~Original partitioning. (b)~Nested dissection reordering into interior chains and reduced system. (c)~Forward and backward passes with communication.}
    \label{fig:two_phase}
    \vspace{-6mm}
\end{figure*}

The precision matrices of the employed spatio-temporal models exhibit a
block-tridiagonal arrowhead (BTA)
sparsity pattern~\cite{rue2005gaussian, dalia_sc}, illustrated in Figure~\ref{fig:two_phase}a.
The matrix is composed of $n$ diagonal blocks arising from a temporal discretization over $n$ time steps.
Each diagonal block, of size $b \times b$, is associated with the discretization of the spatial domain at a given time step using $b$ mesh nodes.
$\bm{D}_i \in \mathbb{R}^{b \times b}$ are the main-diagonal blocks,
$\bm{B}_i \in \mathbb{R}^{b \times b}$ ($i=1,\ldots,n{-}1$) are sub-diagonal blocks encoding temporal dependence between consecutive time steps,
and $\bm{C}_i \in \mathbb{R}^{a \times b}$ together with the arrowhead tip $\bm{T} \in \mathbb{R}^{a \times a}$ arise from $a$ fixed-effect (regression) coefficients that couple to every temporal block through the observation model.
The conditional precision $\bm{Q}_c$ has full BTA structure, while the prior $\bm{Q}_p$ is block-tridiagonal (BT, i.e., $\bm{C}_i = 0$): the arrowhead arises from the observation term $\tau\bm{A}^T\bm{A}$ in $\bm{Q}_c$.
For multivariate models with $k$ correlated response variables, $b = k \cdot n_s$ ($n_s$~spatial locations), increasing both per-block cost and the number of hyperparameters $d = 4k + \binom{k}{2}$ (each variable contributes 4 parameters; each pair adds a coupling parameter).
In the models we benchmark, $n$ ranges from 5 to 512 temporal steps and $b$ from 92 to 13{,}455. The arrowhead width ranges from $a = 2$ to $6$, resulting in precision matrix dimensions $N = b \cdot n + a$.

\subsection{Efficient Block Solver}
\label{sec:bg:solver}

For BTA matrices, the Serinv library~\cite{serinv} provides routines for Cholesky factorization in $O(n \cdot b^3)$ operations, triangular solve, and \emph{selected inversion} (SI), which computes entries of $\bm{Q}^{-1}$ at positions where $\bm{Q}$ is nonzero, also in $O(n \cdot b^3)$. Although the total number of latent parameters $N \approx n \cdot b$ ranges from hundreds to 1.9~million, small by modern machine learning standards, the computational challenge lies in the \emph{per-block density}: each temporal step requires multiple dense $b \times b$ linear algebra routines per optimization step, and the factorization must store $n{-}1$ Schur-complement carries of size $b \times b$.
For the largest models, the carries alone consume 63\,GiB, and together with workspace memory for factorization and triangular solves, require distribution across multiple GPUs (Section~\ref{sec:method:multigpu}).

\subsection{Gradient Computation in INLA}
\label{sec:bg:gradient}

Both stages of the pipeline described in Section~\ref{sec:bg:inla} require differentiating~$f$ through the BTA factorizations above.
DALIA~\cite{dalia_sc}, the state-of-the-art HPC implementation of INLA, has scaled inference to million-parameter models with up to three orders of magnitude speedup over the reference R-INLA implementation~\cite{rue2009approximate, lindgren2011explicit}.
DALIA optimizes~$f$ using an L-BFGS-B~\cite{nocedal2006numerical} algorithm. This requires the evaluation of $f$ and its gradient in every iteration and computes the gradients via central finite differences:
\begin{equation}
    \frac{\partial f}{\partial \theta_j} \approx
    \frac{f(\bm{\theta} + h \bm{e}_j) - f(\bm{\theta} - h \bm{e}_j)}{2h},
    \label{eq:fd}
\end{equation}
where $h$ is a small step size ($h{=}10^{-3}$ for gradients, $h{=}5{\times}10^{-3}$ for the Hessian) and $\bm{e}_j$ is the $j$-th unit vector, so each evaluation perturbs a single hyperparameter. The scheme incurs an inherent truncation error of $O(h^2)$~\cite{nocedal2006numerical} and requires $2d{+}1$ evaluations of~$f$ for $d$ hyperparameters.
Automatic differentiation (AD)~\cite{griewank2008evaluating, baydin2018automatic} computes exact derivatives by applying the chain rule to each elementary operation.
In reverse mode, the program is evaluated forward while recording a trace of operations; a backward pass then propagates derivatives from the output back through this trace (also called the \emph{tape}) to accumulate the gradient.
Because non-linear operations (e.g., Cholesky factorization) cannot be inverted from their output alone, the backward pass requires intermediate results from the forward pass; these can either be stored during the forward pass and provided to the backward pass, or recomputed when needed, leading to a tradeoff between memory and compute.
Unlike forward-mode AD, which scales as $O(d)$, reverse-mode AD yields the full gradient of a scalar objective in a single backward pass at a proven cost of $3$--$5\times$ one forward evaluation~\cite{griewank2008evaluating}, though realized speedups also depend on per-evaluation framework efficiency (Section~\ref{sec:eval:framework}).

Beyond gradient computation, INLA requires the Hessian of~$f$ at the optimum to quantify posterior uncertainty over the hyperparameters. With finite differences, a second-order central difference scheme requires $2d^2{+}1$ function evaluations; Section~\ref{sec:method:hessian} describes how AD reduces this cost.

%% file: sections/method.tex
\section{Structure-Exploiting AD for INLA}
\label{sec:method}

\subsection{Custom Backward Pass for BTA Systems}
\label{sec:method:bta}

Recall that within INLA's objective function $f$ (Eq.~\ref{eq:objective}): $\bm{Q}_c$ has BTA structure (Figure~\ref{fig:two_phase}a), while $\bm{Q}_p$ is block-tridiagonal (BT) without the arrowhead.
The dominant cost is computing $\log|\bm{Q}_c|$, $\log|\bm{Q}_p|$, and $\bm{x}^*$ via Cholesky factorizations and triangular solves.
Naive reverse-mode AD through these operations is impractical: the standard AD rule for differentiating a Cholesky factorization~\cite{murray2016differentiation} requires $\bm{L}^{-1}$, which is dense even when $\bm{L}$ is sparse, degrading backward complexity from $O(n \cdot b^3)$ to $O(N^3)$ where $N = n \cdot b$.
Expressing the factorization as a block-by-block loop avoids the dense inverse, but AD must store all intermediate states across iterations for the backward pass, exceeding GPU memory for large models.
We address this through three techniques: (a)~a fused forward pass that reduces the stored intermediates, (b)~carry-based reconstruction that recovers the per-block components of $\bm{L}$ on the fly during the backward pass from compact stored intermediates, and (c)~a hand-derived analytical gradient that decomposes Eq.~\ref{eq:objective} into independently differentiable terms.

\subsubsection{Fused Forward Pass}
The posterior mode satisfies $\bm{Q}_c\bm{x}^* \!= \! \bm{L}\bm{L}^T\bm{x}^* = \bm{r}$
which is solved for $\bm{x}^*$ using a lower triangular solve $\bm{L}\bm{z} = \bm{r}$ and an upper triangular solve $\bm{L}^T\bm{x}^* = \bm{z}$.
Our fused pass (Algorithm~\ref{alg:fused}) combines the Cholesky factorization with the lower triangular solve in one sweep, treating each diagonal factor $\bm{L}_{D_i}$ as a per-step temporary rather than storing the full factor~$\bm{L}$.

\begin{algorithm}[t]
    \caption{Fused BTA Cholesky with lower triangular solve. Fused steps are shown in \begin{tikzpicture} \fill[fusedcolor] (0,0) rectangle (0.25cm,0.25cm); \end{tikzpicture}.
    }
    \label{alg:fused}
    \begin{algorithmic}[1]
        \Require BTA blocks $\{\bm{D}_i\}_{i=1}^{n}$, $\{\bm{B}_i\}_{i=1}^{n-1}$, $\{\bm{C}_i\}_{i=1}^{n}$ of $\bm{Q}_c$, arrowhead tip $\bm{T}$, RHS $\bm{r}$
        \Ensure Schur carries $\{\bm{S}_i\}_{i=1}^{n-1}$, lower solve vectors $\{\bm{z}_i\}_{i=1}^{n}$, $\bm{z}_T$, arrow factor $\bm{L}_T$
        \State $\bm{S}_0 \gets \bm{0}$, \quad $\tilde{\bm{T}}_0 \gets \bm{T}$, \quad $\bm{C}_0^L \gets \bm{0}$, \quad $\bm{L}_{B_0} \gets \bm{0}$, \quad $\bm{z}_0 \gets \bm{0}$, \quad $\bm{v}_T \gets \bm{0}$
        \For{$i = 1$ to $n$}
            \State $\bm{L}_{D_i} \gets \chol(\bm{D}_i - \bm{S}_{i-1})$
            \State $\bm{L}_{C_i} \gets (\bm{C}_i - \bm{C}_{i-1}^L \bm{L}_{B_{i-1}}^T) \bm{L}_{D_i}^{-T}$
            \State \textcolor{fusedcolor}{$\bm{z}_i \gets \bm{L}_{D_i}^{-1}(\bm{r}_i - \bm{L}_{B_{i-1}} \bm{z}_{i-1})$, \quad $\bm{v}_T \gets \bm{v}_T + \bm{L}_{C_i}\bm{z}_i$}
            \If{$i < n$}
                \State $\bm{L}_{B_i} \gets \bm{B}_i \bm{L}_{D_i}^{-T}$, \quad $\bm{S}_i \gets \bm{L}_{B_i} \bm{L}_{B_i}^T$
            \EndIf
            \State $\tilde{\bm{T}}_i \gets \tilde{\bm{T}}_{i-1} - \bm{L}_{C_i}\bm{L}_{C_i}^T$, \quad $\bm{C}_i^L \gets \bm{L}_{C_i}$
        \EndFor
        \State $\bm{L}_T \gets \chol(\tilde{\bm{T}}_n)$, \quad \textcolor{fusedcolor}{$\bm{z}_T \gets \bm{L}_T^{-1}(\bm{r}_T - \bm{v}_T)$}
    \end{algorithmic}
    
\end{algorithm}

Because of this fusion, $\bm{L}_{D_i}$ and $\bm{L}_{B_i}$ are only needed within step~$i$ and are discarded after computing the Schur-complement carry $\bm{S}_i$ (the $b {\times} b$ matrix that propagates coupling from block~$i$ to block~$i{+}1$) and the lower triangular solve vector $\bm{z}_i$.
Only $\{\bm{S}_i\}$ and $\{\bm{z}_i\}$ are retained, halving storage relative to the full factor ($n$ vs ${\sim}2n$ dense $b{\times}b$ blocks).

\subsubsection{Upper Triangular Solve and L Reconstruction}
The second stage $\bm{L}^T\bm{x}^* = \bm{z}$ sweeps $i$ from block~$n$ to~$1$:
\begin{align}
    \bm{x}_T^* &= \bm{L}_T^{-T} \bm{z}_T \label{eq:utsub_arrow} \\
    \bm{x}_n^* &= \bm{L}_{D_n}^{-T}(\bm{z}_n - \bm{L}_{C_n}^T \bm{x}_T^*) \label{eq:utsub_init} \\
    \bm{x}_i^* &= \bm{L}_{D_i}^{-T}(\bm{z}_i - \bm{L}_{B_i}^T \bm{x}_{i+1}^* - \bm{L}_{C_i}^T \bm{x}_T^*) \label{eq:utsub_step}
\end{align}
This, together with the selected inversion (SI; Phase~A below) and gradient accumulations (Phases~A--C), requires $\bm{L}_{D_i}$ at each block and $\bm{L}_{B_i}$ for $i < n$.
Rather than storing them, we reconstruct on the fly: at block~$i$, the stored carry $\bm{S}_{i-1}$ together with the input blocks $\bm{D}_i$ and $\bm{B}_i$ suffice to recompute $\bm{L}_{D_i}$ via one Cholesky factorization and two triangular solves.
This trades compute for memory: one additional Cholesky per block, with peak working memory of $O(b^2)$.
All backward operations share the same sweep direction and reconstructed factors, so they are fused into a single pass over the blocks.

\subsubsection{Analytical Gradient Decomposition}
The objective $f(\bm{\theta})$ depends on~$\bm{\theta}$ both directly (through $\bm{Q}_p$, $\bm{Q}_c$, and the priors) and indirectly through the posterior mode $\bm{x}^*(\bm{\theta}) = \bm{Q}_c^{-1}\bm{r}$.
By the chain rule, the total derivative is $\frac{df}{d\theta_k} = \frac{\partial f}{\partial \theta_k} + \frac{\partial f}{\partial \bm{x}^{*T}} \frac{d\bm{x}^*}{d\theta_k}$.
The second term vanishes by the envelope theorem~\cite{milgrom2002envelope}: $\bm{x}^*$ maximizes $f$ with respect to~$\bm{x}$, so the stationarity condition $\partial f/\partial \bm{x} = \bm{0}$ holds at $\bm{x}{=}\bm{x}^*$, and $\bm{x}^*$ can be treated as a constant when differentiating with respect to~$\bm{\theta}$.
Differentiating each term of Eq.~\ref{eq:objective} with $\bm{x}^*$ held constant gives three independent gradient contributions, detailed below:

\vspace{-4mm}
\begin{multline}
    \frac{\partial f}{\partial \theta_k} = \underbrace{-\frac{1}{2}\tr\!\left(\bm{Q}_c^{-1}\frac{\partial \bm{Q}_c}{\partial \theta_k}\right)}_{\text{Phase A}} - \underbrace{\frac{1}{2}\bm{x}^{*T}\frac{\partial \bm{Q}_p}{\partial \theta_k}\bm{x}^*}_{\text{Phase B}} \\
    \underbrace{+ \frac{1}{2}\tr\!\left(\bm{Q}_p^{-1}\frac{\partial \bm{Q}_p}{\partial \theta_k}\right)}_{\text{Phase C}} + \frac{\partial \log p}{\partial \theta_k} + \frac{\partial}{\partial \theta_k}\log \ell(\bm{y}\mid\bm{x}^*,\bm{\theta})
    \label{eq:grad_decomp}
\end{multline}

Each phase is computed analytically from the stored carries $\{\bm{S}_i\}$ and vectors $\{\bm{z}_i\}$, requiring $O(n \cdot b^3)$ work with $O(b^2)$ working memory per block.
Each factorization produces $n$ carries of size $b \times b$. Since the $\bm{Q}_c$ and $\bm{Q}_p$ sweeps run sequentially, only one set is live at a time, giving a peak memory footprint of $n \cdot b^2$ (carries) plus $O(b^2)$ per-block workspace for the reconstructed factors, independent of the number of intermediates that naive AD would store.

\paragraph{Phase~A - Selected inversion gradients for $\log|\bm{Q}_c|$}
The gradient of a log-determinant is $\frac{\partial \log|\bm{Q}|}{\partial \theta_k} = \tr\!\left(\bm{Q}^{-1}\frac{\partial \bm{Q}}{\partial \theta_k}\right)$~\cite{petersen2012matrix, rue2005gaussian}.
Naively, this requires the full inverse $\bm{Q}^{-1}$, which is dense and costs $O(N^3)$ to compute.
However, expanding the trace of a product gives $\tr(\bm{A}\bm{B}) = \sum_{ij} A_{ij}B_{ji}$, which reduces to an element-wise sum without forming the full product $\bm{A}\bm{B}$.
Since both $\bm{Q}^{-1}$ and $\frac{\partial \bm{Q}}{\partial \theta_k}$ are symmetric and the latter has the same BTA sparsity pattern as~$\bm{Q}$, every term with a zero entry in $\frac{\partial \bm{Q}}{\partial \theta_k}$ vanishes, and we only need $\bm{Q}^{-1}$ at the nonzero positions of~$\bm{Q}$: the block-diagonal $\bm{Z}_{D_i} = [\bm{Q}^{-1}]_{ii}$, block-sub-diagonal $\bm{Z}_{B_i} = [\bm{Q}^{-1}]_{i+1,i}$, arrowhead column $\bm{Z}_{C_i} = [\bm{Q}^{-1}]_{T,i}$, and arrowhead tip $\bm{Z}_T = [\bm{Q}^{-1}]_{T,T}$. These are precisely the entries computed by \emph{Selected Inversion}~\cite{serinv} which sweeps backward from block~$n$ to~$1$, recovering $\bm{Z}_{D_i}$, $\bm{Z}_{B_i}$, $\bm{Z}_{C_i}$, and $\bm{Z}_T$ from the Cholesky factors $\bm{L}_{D_i}$, $\bm{L}_{B_i}$, $\bm{L}_{C_i}$, and $\bm{L}_T$.

The block Jacobians $\frac{\partial \bm{D}_i}{\partial \theta_k}$, $\frac{\partial \bm{B}_i}{\partial \theta_k}$, etc.\ are inexpensive to obtain from the closed-form block construction.
The full trace then reduces to a sum of per-block traces:
\begin{multline}
    \frac{\partial \log|\bm{Q}_c|}{\partial \theta_k} = \sum_{i=1}^{n} \tr\!\left(\bm{Z}_{D_i}\frac{\partial \bm{D}_i}{\partial \theta_k}\right) + 2\sum_{i=1}^{n-1}\tr\!\left(\bm{Z}_{B_i}^T\frac{\partial \bm{B}_i}{\partial \theta_k}\right) \\
    {} + 2\sum_{i=1}^{n}\tr\!\left(\bm{Z}_{C_i}^T\frac{\partial \bm{C}_i}{\partial \theta_k}\right) + \tr\!\left(\bm{Z}_T\frac{\partial \bm{T}}{\partial \theta_k}\right)
    \label{eq:si_grad}
\end{multline}
We fuse this trace accumulation into the SI sweep: at each block~$i$, we compute $\bm{Z}_{D_i}$ and $\bm{Z}_{B_i}$, immediately accumulate their gradient contributions, and discard them, so only two blocks of $\bm{Q}^{-1}$ ($\bm{Z}_{D_i}$ and $\bm{Z}_{D_{i+1}}$) are ever live simultaneously.

\paragraph{Phase~B - Quadratic form gradients}
The quadratic form $\bm{x}^{*T}\bm{Q}_p\bm{x}^*$ expands via the BT block structure as:
\vspace{-2mm}
\begin{equation}
    \bm{x}^{*T}\bm{Q}_p\bm{x}^* = \sum_{i=1}^{n} \bm{x}_i^{*T} \bm{D}_i \bm{x}_i^* + 2\sum_{i=1}^{n-1} \bm{x}_i^{*T} \bm{B}_i^{T} \bm{x}_{i+1}^*
    \label{eq:quad_expand}
    \vspace{-1mm}
\end{equation}
Since $\bm{x}^*$ is treated as fixed at the mode, differentiating with respect to $\theta_k$ passes through to the blocks:
\vspace{-2mm}
\begin{equation}
    \hspace{-3mm}\frac{\partial}{\partial \theta_k}(\bm{x}^{*T} \bm{Q}_p\bm{x}^*) = \sum_{i=1}^{n} \bm{x}_i^{*T} \frac{\partial \bm{D}_i}{\partial \theta_k}\bm{x}_i^* + 2\sum_{i=1}^{n-1}\bm{x}_i^{*T} \frac{\partial \bm{B}_i}{\partial \theta_k}\bm{x}_{i+1}^*
    \label{eq:quad_grad}
    \vspace{-1mm}
\end{equation}
This requires only the already-computed posterior mode~$\bm{x}^*$ and the block Jacobians of~$\bm{Q}_p$.

\paragraph{Phase~C - Prior log-determinant gradients}
The term $\log|\bm{Q}_p|$ has the structure of $\log|\bm{Q}_c|$ but uses BT factorization (no arrowhead), so the same carry-based technique applies.

While we instantiate this approach for BTA matrices, the underlying pattern generalizes to any block recurrence where (i)~a compact carry state summarizes all prior blocks and (ii)~per-block factors can be reconstructed from that carry and the local input data.
This includes block-tridiagonal, block-banded, and related elimination-based factorizations.

\subsection{Extension to Multivariate Models}
\label{sec:method:coreg}

The previous section developed the backward pass for a single response variable. We now extend it to multivariate spatio-temporal models that jointly model $k$ correlated response variables (e.g., three correlated pollutants in air quality monitoring).
The BTA precision matrix retains the same block-tridiagonal pattern, but each diagonal block~$\bm{D}_t$ now groups all $k$ variables at a given time step into a \emph{super-block} of size $b = k \cdot n_s$, where $n_s$ is the number of spatial locations.
This super-block is itself a $k \times k$ grid of $n_s \times n_s$ sub-blocks, where each sub-block encodes the spatial coupling between a pair of variables.
Let $\bm{Q}_m(t) \in \mathbb{R}^{n_s \times n_s}$ denote the spatial precision matrix of the $m$-th variable at time step~$t$, as if that variable were modeled independently.
Each $(i,j)$ sub-block of the super-block at time~$t$ is a weighted combination of these per-variable matrices:

\vspace{-4mm}
\begin{equation}
    \bm{Q}_{\text{super}}^{(ij)}(t) = \sum_{m=1}^{k} W_{ij}^{(m)}(\bm{\sigma}, \bm{\lambda}) \cdot \bm{Q}_m(t)
    \label{eq:coreg_superblock}
    \vspace{-3mm}
\end{equation}

where $W_{ij}^{(m)}$ are entries of the coregional weight matrix, derived from the linear model of coregionalization~\cite{schmidt2003bayesian}, parameterized by scale $\bm{\sigma}$ and coupling parameters $\bm{\lambda}$, both part of the hyperparameter vector~$\bm{\theta}$, for details see~\cite{dalia_sc}

The forward factorization (Section~\ref{sec:method:bta}) operates on these super-blocks unchanged.
We formulate the super-block as Eq.~\ref{eq:coreg_superblock} specifically to exploit the fact that each hyperparameter in~$\bm{\theta}$ affects exactly one of the two factors, reducing per-step work in the backward pass.
A per-variable hyperparameter $\theta_k$ (e.g., spatial range) affects~$\bm{Q}_m$ but not the weights:
\begin{equation}
    \frac{\partial \bm{Q}_{\text{super}}^{(ij)}}{\partial \theta_k} = W_{ij}^{(m)} \cdot \frac{\partial \bm{Q}_m}{\partial \theta_k}
    \label{eq:coreg_jac_variate}
\end{equation}
A coupling hyperparameter $\theta_k$ (i.e., one of the~$\sigma_j$ or~$\lambda_j$) affects the weights but not~$\bm{Q}_m$:
\begin{equation}
    \frac{\partial \bm{Q}_{\text{super}}^{(ij)}}{\partial \theta_k} = \sum_{m=1}^{k} \frac{\partial W_{ij}^{(m)}}{\partial \theta_k} \cdot \bm{Q}_m
    \label{eq:coreg_jac_coupling}
\end{equation}
Both enter the SI gradient trace (Eq.~\ref{eq:si_grad}) through the block Jacobians $\frac{\partial \bm{D}_i}{\partial \theta_k}$.
The per-variable case (Eq.~\ref{eq:coreg_jac_variate}) requires the weight $W_{ij}^{(m)}$ and the per-model derivative $\frac{\partial \bm{Q}_m}{\partial \theta_k}$; the coupling case (Eq.~\ref{eq:coreg_jac_coupling}) requires all~$k$ per-model blocks $\bm{Q}_m(t_i)$ themselves.
In practice, the $\bm{Q}_m$ are never formed as separate matrices; they share the same underlying spatial components and are combined with the coregionalization weights on the fly during super-block assembly.
Since the per-model blocks and Jacobians are reconstructed from the same stored carries and discarded after each step, this adds no persistent memory beyond the existing $O(b^2)$ per-block workspace.

\subsection{Multi-GPU Distributed AD}
\label{sec:method:multigpu}

The BTA factorization is inherently block-sequential, where the computation related to each diagonal block depends on the Schur-complement carry from the previous diagonal block.
The primary motivation for distribution is \emph{memory}: for the largest models, the stored carries ($n \cdot b^2$ elements) could exceed the capacity of a single node (GPU and host combined), making distribution a necessity to run these models at all.
We adopt Serinv's two-phase domain decomposition~\cite{serinv} for the forward factorization and extend it to the backward pass, distributing both across~$P$ GPUs. On top of enabling larger models, this algorithm also exposes concurrency by reordering the blocks such that the majority of the work (Phase~1) can be done in parallel with no communication, while the remaining work (Phase~2) requires only a single round of communication and redundant local solves.
The hand-derived backward pass from Section~\ref{sec:method:bta} is essential here: it provides explicit block-by-block control that allows inserting MPI communication and staging carries to CPU memory, which would not be trivially possible within an automatic AD tape.

\paragraph{Two-Phase Partitioning}
The $n$ temporal blocks are divided into $P$ contiguous partitions (Figure~\ref{fig:two_phase}a).
The blocks at partition interfaces (\emph{boundary} blocks) couple to blocks in both adjacent partitions through the Schur complement carry; the remaining \emph{interior} blocks couple only within their own partition.
Serinv's nested dissection reordering~\cite{serinv} (Figure~\ref{fig:two_phase}b, regions delimited by red lines) separates these two groups: interior blocks stay on their assigned GPU as a self-contained chain, while boundary blocks are collected into a small shared \emph{reduced system} of $2P{-}1$ blocks that captures all cross-partition coupling.
For example, with $n{=}6$ and $P{=}2$: GPU~0 keeps $\bm{D}_1, \bm{D}_2$ as its interior chain, GPU~1 keeps $\bm{D}_5$, and the boundary blocks $\bm{D}_3, \bm{D}_4, \bm{D}_6$ form the reduced system.
In Phase~1 (Figure~\ref{fig:two_phase}c), all GPUs factorize their interior chains concurrently: GPU~0 processes $\bm{D}_1 \to \bm{D}_2$, GPU~1 processes $\bm{D}_5$.
In Phase~2, an MPI allgather exchanges boundary data, and each GPU redundantly solves the reduced system ($\bm{D}_3, \bm{D}_4, \bm{D}_6$) to avoid communication.
For the backward pass, the order reverses: Phase~2$'$ solves the reduced system first to produce boundary selected inversion values, then Phase~1$'$ runs concurrent backward sweeps seeded by these values, followed by an allreduce of the $d$ gradient scalars.

\paragraph{Communication Pattern}
Phase~1 requires no inter-rank communication.
After Phase~1, MPI allgathers collect the $2P{-}1$ boundary diagonal ($b \times b$), lower ($b \times b$), and arrow blocks ($a \times b$), plus the lower triangular solve RHS vectors ($b \times 1$), from all ranks; a negligible allreduce sums the arrow tip ($a \times a$) and the interior log-determinant (scalar).
In Phase~2, each rank assembles and factorizes the reduced system ($2P{-}1$ blocks) locally.
A final MPI allreduce of $d$ gradient scalars combines per-rank contributions.
The total number of distinct elements exchanged across all collectives is $(4P{-}2)b^2 + (2P{-}1)(a{+}1)b + a^2 + d + 1$, dominated by the $O(Pb^2)$ boundary blocks.

\paragraph{CPU-Staged Schur Carries}
The Schur carries ($n_{\text{local}} \times b^2$ elements per rank) are written to CPU memory during the forward pass and read back to GPU one block at a time during the backward pass for L reconstruction.
This bounds per-GPU memory to $O(b^2)$ working memory regardless of $n_{\text{local}}$, at the cost of $n_{\text{local}}$ CPU$\leftrightarrow$GPU transfers.
The same staging is applied to the reduced-system factorization when the $2P{-}1$ blocks cannot be held simultaneously on GPU, keeping the $O(b^2)$ bound throughout.

\subsection{Posterior Hessian via AD Gradients}
\label{sec:method:hessian}

Post optimization, INLA computes the Hessian $\bm{H} = \nabla^2 f(\bm{\theta}^*)$ at the optimum to approximate the posterior uncertainty of the hyperparameters.
Rather than $2d^2{+}1$ objective function evaluations (second-order FD approach), we compute this Hessian via first-order finite differences of AD gradients:
\begin{equation}
    \bm{H}_{:,j} \approx \frac{\nabla f(\bm{\theta}^* {+} h\bm{e}_j) - \nabla f(\bm{\theta}^* {-} h\bm{e}_j)}{2h}
    \label{eq:hessian_fd_grad}
\end{equation}
This reduces the cost from $2d^2{+}1$ objective evaluations to $2d+1$ calls to the AD gradient routine.
Because the Hessian computation is a one-time post-optimization step, fully second-order AD would add complexity for a modest gain: the Hessian differentiates the $d$-vector gradient, not a scalar function, so each of its $d$ columns still requires a separate reverse pass.

\subsection{Implementation}
\label{sec:method:impl}

Our open-source implementation, ADELIA, extends the DALIA framework~\cite{dalia_sc} with a JAX backend.
ADELIA coexists with DALIA's original FD implementation, which uses CuPy with Serinv's~\cite{serinv} GPU-optimized BTA solver, allowing direct comparison on identical model configurations.
JAX's \texttt{jax.custom\_vjp} primitive registers our structure-exploiting backward pass for the BTA factorization, while JAX's standard reverse-mode AD handles the remaining terms (prior log-probabilities, likelihood); \texttt{jax.jit} compiles each stage via XLA~\cite{jax2018github} for GPU execution.
Multi-GPU communication uses mpi4jax~\cite{mpi4jax} for JAX-compatible MPI operations within XLA-compiled code.

%% file: sections/evaluation.tex
\section{Experimental Evaluation}
\label{sec:evaluation}
We organize the evaluation around eight questions a practitioner would follow when adopting AD for structured sparse matrix computations.
We first establish that structure-exploiting AD is necessary, not optional~(\S\ref{sec:eval:structure}), then validate gradient correctness and show that AD gradients enable convergence where FD stalls~(\S\ref{sec:eval:gradval}).
We measure per-gradient speedup over FD on equal hardware~(\S\ref{sec:eval:minres}), translate these to end-to-end wall-clock gains~(\S\ref{sec:eval:wallclock}), and ask whether the advantage generalizes across problem sizes~(\S\ref{sec:eval:scaling}).
We then disentangle framework effects from algorithmic gains~(\S\ref{sec:eval:framework}), examine what it costs FD in number of GPUs and energy to compensate AD speed~(\S\ref{sec:eval:efficiency}), and characterize where time and memory concentrate inside the gradient pipeline~(\S\ref{sec:eval:breakdown}).
\vspace{-2mm}
\subsection{Experimental Setup}
\label{sec:eval:setup}
\vspace{-3mm}
\paragraph{Benchmark Models}
\begin{table}[h]
    \caption{Benchmark model characteristics. $P_{\min}$: minimum number of GPUs to fit the model.}
    \label{tab:models}
    \centering
    \small
    \setlength{\tabcolsep}{4pt}
    \begin{tabular}{llrrrrr}
        \toprule
        Model & Application & Latent & $d$ & $b$ & $n$ & $P_{\min}$ \\
        \midrule
        GST-S  & Synthetic   & 466     & 4  & 92     & 5   & 1 \\
        GST-M  & Synthetic   & 81K     & 4  & 812    & 100 & 1 \\
        GST-L  & Synthetic   & 1.0M    & 4  & 4,002  & 250 & 1 \\
        GST-T  & Temperature & 1.0M    & 4  & 2,865  & 365 & 1 \\
        GST-C2 & Synthetic   & 8.5K    & 9  & 708    & 12  & 1 \\
        GST-C3 & Synthetic   & 8.5K    & 15 & 1,062  & 8   & 1 \\
        \midrule
        SA1    & Benchmark   & 964K    & 15 & 5,019  & 192 & 1 \\
        AP1    & Air poll.   & 606K    & 15 & 12,630 & 48  & 4 \\
        WA1    & Benchmark   & 1.9M    & 15 & 3,741  & 512 & 4 \\
        WA2    & Benchmark   & 646K    & 15 & 13,455 & 48  & 4 \\
        \bottomrule
    \end{tabular}
    \vspace{-3mm}
\end{table}
GST-S/M/L (Table~\ref{tab:models}) are univariate synthetic benchmarks of increasing size; GST-T is a real-world temperature dataset~\cite{gaedkeIntegrated2024} with daily observations at 2{,}865 weather stations over one year ($n_t{=}365$).
GST-C2 and GST-C3 add multivariate (coregionalization) structure with $k{=}2$ and $k{=}3$ variables, respectively~\cite{dalia_sc}.
AP1 is a real-world air pollution model that estimates particulate matter (PM2.5, PM10) and ozone (O$_3$) concentrations over northern Italy~\cite{dalia_sc}.
SA1, WA1, and WA2 are production-scale benchmarks from DALIA~\cite{dalia_sc} that exercise different scaling regimes: SA1 has moderate blocks over many time steps ($n_t{=}192$), WA1 stresses temporal depth ($n_t{=}512$), and WA2 stresses spatial density ($n_s{=}4{,}485$).
All four are trivariate models with $d{=}15$ hyperparameters controlling spatial range, temporal smoothness, and cross-variable coupling.

\paragraph{Hardware}
All experiments run on the CSCS Alps supercomputer~\cite{10.1145/3757348.3757365} using NVIDIA GH200 Grace Hopper nodes, each with four Hopper-based GPUs (96\,GiB HBM3 each) and a 72-core ARM Neoverse V2 CPU (Grace) with 480\,GiB LPDDR5X memory, connected via NVLink-C2C. All four GPUs per node are utilized; multi-node runs communicate over an HPE Slingshot-11 interconnect.

\paragraph{Software}
We use ADELIA (extending DALIA~\cite{dalia_sc}) with JAX 0.8.1 for AD and mpi4jax for distributed communication; DALIA with CuPy 13.6.0 provides baseline FD operations. MPI uses Cray MPICH with CPU-staged transfers (no CUDA-aware MPI or NCCL). We report absolute performance alongside speedups, average over at least 10 runs after warmup, and show 95\% confidence intervals~\cite{hoefler2015scientific}.
JIT compilation is a one-time cost (39--246\,s across models), amortized over long optimization runs.

\subsection{Structure-Exploiting Differentiation}
\label{sec:eval:structure}

\researchquestion{Is the custom backward pass necessary, or could generic AD approaches work?}

\smallskip
We evaluate five strategies  on a single GPU (Table~\ref{tab:structure_comparison}):
\textbf{FD} (central finite differences, $2d{+}1$ evaluations per gradient);
\textbf{AD-Dense} (dense Cholesky on the full $N{\times}N$ matrix, storing the dense factor);
\textbf{AD-Loop} (BTA Cholesky as a \texttt{lax.scan} loop, storing all $n$ loop carries and per-step $L$ factors in the AD tape);
\textbf{AD-Ckpt} (gradient checkpointing~\cite{chen2016training}, trading recomputation for reduced tape storage);
and \textbf{ADELIA} (storing only the $n$ Schur carries and reconstructing $L$ factors on the fly).

\begin{table}[t]
    \caption{Per-gradient time (s) / peak GPU memory (GiB) for five differentiation strategies. OOM = exceeded GPU memory. Best AD time in \textbf{bold}.}
    \label{tab:structure_comparison}
    \centering
    \footnotesize
    \renewcommand{\arraystretch}{1.15}%
    \setlength{\tabcolsep}{3pt}%
    \begin{tabular}{l ccccc}
        \toprule
        Model & FD & AD-Dense & AD-Loop & AD-Ckpt & ADELIA \\
        \midrule
        GST-S  & .34\,/\,.4  & \textbf{.004\,/\,${<}$.1} & .007\,/\,${<}$.1 & .007\,/\,${<}$.1 & .006\,/\,${<}$.1 \\
        GST-C2 & 1.6\,/\,.2  & .29\,/\,2.9  & \textbf{.075\,/\,.9}  & .085\,/\,1.0  & .079\,/\,.2  \\
        GST-C3 & 2.9\,/\,.2  & .39\,/\,3.3  & \textbf{.083\,/\,1.3}  & .094\,/\,1.5  & .096\,/\,.3  \\
        GST-M  & 2.7\,/\,2.4  & OOM & .78\,/\,8.4  & .87\,/\,6.4  & \textbf{.70\,/\,1.1}  \\
        GST-L  & 56.6\,/\,76.9 & OOM & OOM & OOM & \textbf{23.8\,/\,63.3} \\
        GST-T  & 45.5\,/\,61.2 & OOM & OOM & OOM & \textbf{17.1\,/\,46.5} \\
        \bottomrule
    \end{tabular}
    \vspace{-6mm}
\end{table}

FD scales to all sizes, but is the slowest strategy.
AD-Dense OOMs at GST-M (81K latent variables); AD-Loop and AD-Ckpt OOM at GST-L and GST-T, where the stored states exceed GPU memory.
Only ADELIA fits the million-variable models (GST-L within 63.3\,GiB, GST-T within 46.5\,GiB) while running $2.4\times$ and $2.7\times$ faster than FD, respectively.
Structure-exploiting differentiation is thus a \emph{necessity}, not merely an optimization: every generic AD strategy fails at increasing problem scale.

\subsection{Gradient Correctness and Convergence Quality}
\label{sec:eval:gradval}

\researchquestion{Are the AD gradients numerically correct, and do they lead to high-quality convergence on real-world models?}

Beyond the mathematical derivation in Section~\ref{sec:method}, we validate correctness empirically and evaluate convergence quality.

\paragraph{Exact reference}
On models small enough for AD-Loop (Table~\ref{tab:structure_comparison}), we use it as a reference since it relies on JAX's built-in AD with no custom derivatives. ADELIA matches it: worst-case relative error $1.2 \times 10^{-7}$ (GST-C3); smaller models to $10^{-12}$ or better. To validate the distributed two-phase implementation, we additionally compare to the AD-Loop reference on GST-C3 with P$=$2, achieving $6.8 \times 10^{-6}$.

\paragraph{Finite-difference comparison}
On large models, we validate against central finite differences ($h = 10^{-3}$).
All four distributed models achieve element-wise relative errors of $5 \times 10^{-4}$ to $6 \times 10^{-3}$, consistent with FD truncation $O(h^2)$.

\subsubsection{Convergence case study}
\begin{figure}[t]
    \centering
    \includegraphics[width=\columnwidth]{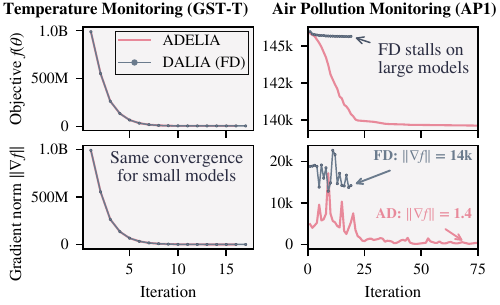}
    \caption{L-BFGS convergence on two real-world models: objective (top) and gradient norm (bottom).}
    \label{fig:convergence_ap1}
    \vspace{-8mm}
\end{figure}

Figure~\ref{fig:convergence_ap1} compares L-BFGS convergence using ADELIA and DALIA from the same initial~$\bm{\theta}$ on two real-world models.
The magnitude of the objective~$f$ varies across models; the goal is to minimize~$f$, not for it to reach zero.
On the smaller models (GST-S through GST-T; GST-T shown left), both methods converge to the same minimum.
As model complexity grows, exact gradients become essential.
On the largest distributed models the gap is significant: on AP1 (shown right), ADELIA converges in 165~iterations, reducing $\|\nabla f\|$ from $4{,}874$ to $1.4$; DALIA's gradient norm is elevated because $\|\nabla f\|$ amplifies per-component FD errors across $d{=}15$ parameters, stalling after 19~iterations at $f{=}145\text{,}660$ vs.\ ADELIA's $f{=}139\text{,}618$.
On SA1, FD stalls at $\|\nabla f\|{=}4{,}019$ with a non-positive-definite Hessian, while ADELIA reaches $\|\nabla f\|{=}1.5$, a $2{,}700\times$ reduction, with a well-conditioned Hessian for posterior uncertainty quantification.
WA1 and WA2 follow suit, with DALIA stalling after 2--5 iterations and ADELIA reaching $10^{4}\times$ lower gradient norms. At this scale, \textbf{ADELIA's exact gradients are not just an optimization but a necessity}: without them, the optimizer cannot reach a valid minimum.

\subsection{Minimum Resources Comparison}
\label{sec:eval:minres}

\researchquestion{With the minimum number of GPUs required to run a model, how much speedup does AD provide over FD?}
\begin{figure*}[t]
    \centering
    \includegraphics[width=\textwidth]{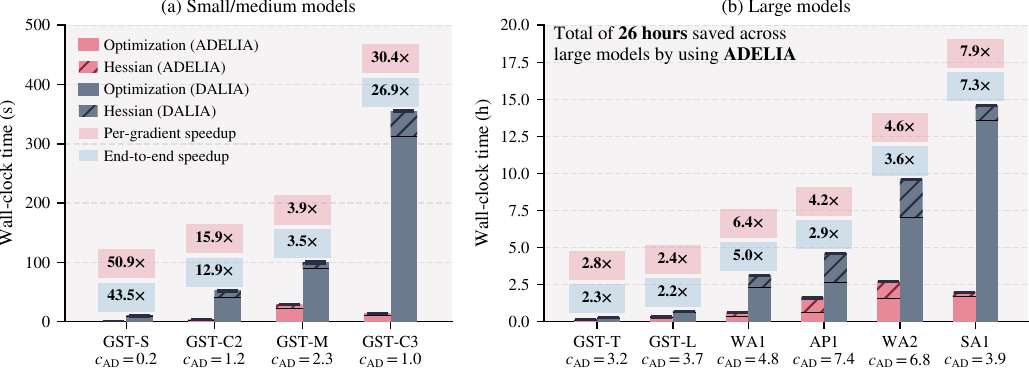}
    \vspace{-6mm}
    \caption{End-to-end wall-clock breakdown (optimization + Hessian) on $P_{\min}$ GPUs, adjusted to the same number of iterations (using ADELIA as baseline) to isolate computational cost from convergence differences.}
    \label{fig:wallclock}
    \vspace{-3mm}
\end{figure*}
\smallskip

We run each model on its minimum feasible GPU count ($P_{\min}$ in Table~\ref{tab:models}) and compare per-gradient times. Figure~\ref{fig:wallclock} presents the results: AD achieves per-gradient speedups of $2.4$--$50.9\times$ across all ten models.
The speedup is bounded by $2d{+}1$ (the FD evaluation count), but the AD backward pass adds overhead, and per-evaluation framework efficiency play a role (Section~\ref{sec:eval:framework}).
We capture this with the \emph{cost ratio} $c_{\mathrm{AD}} = T_{\mathrm{AD}} / t_{\mathrm{eval}}$, which measures how expensive one AD gradient is relative to one FD evaluation ($t_{\mathrm{eval}} = T_{\mathrm{FD}}/(2d{+}1)$).
The observed speedup is then $(2d{+}1)/c_{\mathrm{AD}}$: when $c_{\mathrm{AD}} < 1$, AD gains a per-evaluation \emph{bonus} on top of the count reduction; when $c_{\mathrm{AD}} > 1$, it pays a \emph{penalty} that partially offsets it.
Three regimes emerge.

\paragraph{Small, fully fused models (GST-S, GST-C2, GST-C3)}
XLA compiles the entire forward-backward computation into a single fused kernel, eliminating per-kernel launch overhead and bringing the full AD gradient close to or below the cost of a single CuPy-based FD evaluation ($c_{\mathrm{AD}} \lesssim 1$), a framework advantage quantified in Section~\ref{sec:eval:framework}.

\paragraph{Medium, compute-bound models (GST-M, GST-L, GST-T, SA1)}
The backward pass costs roughly $1.3$--$2.4\times$ the forward, placing the total AD gradient at $2.3$--$3.4\times$ a single forward evaluation, well within the theoretical $3$--$5\times$ bound on the backward-to-forward ratio~\cite{griewank2008evaluating}.

\paragraph{Distributed models (AP1, WA1, WA2)}
These models require 4~GPUs to fit in memory;
Per-block JIT dispatch, CPU$\leftrightarrow$GPU transfers for CPU-staged carries, and the two-phase reduced system overhead add framework cost, raising $c_{\mathrm{AD}}$ to $4.8$--$7.4$.
Despite this penalty, the $2d{+}1{=}31$ evaluation count advantage still yields $4.2$--$6.4\times$ speedups.

\subsection{End-to-End Wall-Clock}
\label{sec:eval:wallclock}

\researchquestion{How do per-gradient speedups translate to end-to-end wall-clock time?}

\smallskip
Figure~\ref{fig:wallclock} decomposes the INLA run into optimization and posterior Hessian computation.
As shown in Section~\ref{sec:eval:gradval}, FD and AD can follow different optimization trajectories.
To isolate computational cost from convergence quality, our end-to-end comparison uses a fixed-work budget: we run ADELIA until convergence for each model, and use the resulting iteration count for both methods.

The Hessian stage (Section~\ref{sec:method:hessian}) adds a second advantage: AD computes each Hessian column via two gradient evaluations, while FD requires $O(d^2)$ objective function evaluations via second-order finite differences, yielding $1.1$--$23.3\times$ Hessian speedups.
Overall end-to-end speedups range from $2.2\times$ (GST-L) to $7.3\times$ (SA1) for the production-scale models, with distributed models at $2.9$--$5.0\times$.
In absolute terms, AD reduces the projected INLA run from 14.6\,h to 2.0\,h for SA1 and from 9.6\,h to 2.7\,h for WA2 on equal hardware.

\subsection{Problem Size Impact Study}
\label{sec:eval:scaling}

\researchquestion{Does AD's advantage hold consistently across problem sizes, or is it model-specific?}

\smallskip
We study how speedup scales with problem size along two axes using the WA1 and WA2 model families on a single GPU. The largest sizes here are smaller than the production configurations in Table~\ref{tab:models}, which require multiple GPUs; this study isolates the scaling trend from distribution overhead.

\begin{figure*}[t]
    \centering
    \includegraphics[width=\textwidth]{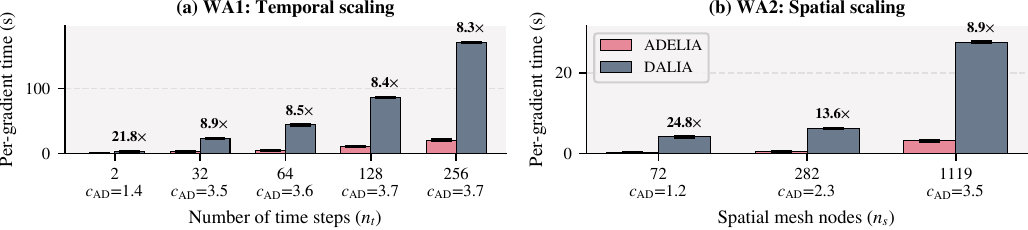}
    \vspace{-6mm}
    \caption{AD speedup vs.\ problem size.
    (a)~WA1 with increasing temporal resolution ($n_t = 2$ to $256$, $n_s{=}1{,}247$).
    (b)~WA2 with increasing spatial resolution ($n_s = 72$ to $1{,}119$, $n_t{=}48$).
    Speedup annotations on FD bars; $c_{\mathrm{AD}}$ values below each size label.}
    \label{fig:scaling}
    \vspace{-5mm}
\end{figure*}

For WA1 temporal scaling (Figure~\ref{fig:scaling}a), $c_{\mathrm{AD}}$ rises from $1.4$ to $3.7$, converging quickly.
At small~$n$, the backward pass costs $1.6\times$ the forward; at large~$n$, the loop is compute-bound ($\beta = 2.2$), pushing $c_{\mathrm{AD}}$ to its asymptote.

Figure~\ref{fig:scaling}(b) shows the same pattern for WA2 spatial scaling ($n_s{=}72$ to $1{,}119$): $c_{\mathrm{AD}}$ rises from $1.2$ to $3.5$ as $O(b^3)$ computation overtakes kernel-launch overhead.
Both axes converge to $c_{\mathrm{AD}} \approx 3.5$, confirming that the asymptotic cost ratio is independent of whether the problem grows in time or space. AD retains $8$--$25\times$ speedups across the full range.

\subsection{Framework Effect Decomposition}
\label{sec:eval:framework}

\researchquestion{How much of the observed speedup comes from XLA compilation efficiency versus the algorithmic advantage of AD?}

\smallskip
ADELIA uses JAX/XLA, while DALIA's FD baseline uses CuPy with Serinv~\cite{serinv}.
We isolate the framework effect by running the same evaluation on both stacks, measuring $r = t_{\mathrm{Serinv}} / t_{\mathrm{JAX}}$.
At framework parity ($r{=}1$), the speedup would be purely algorithmic: $(2d{+}1)/(1{+}\beta)$, where $\beta = t_{\mathrm{bwd}}/t_{\mathrm{fwd}}$ is the backward-to-forward cost ratio.
\begin{figure}[t]
    \centering
    \includegraphics[width=\linewidth]{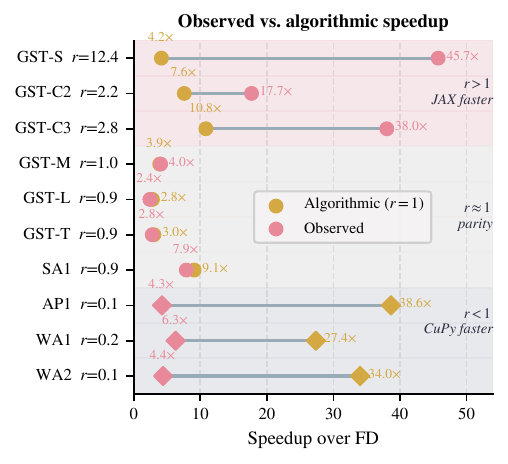}
    \vspace{-7mm}
    \caption{Framework effect decomposition: observed speedup (pink) vs.\ algorithmic speedup at framework parity (gold).}
    \label{fig:framework}
    \vspace{-6.6mm}
\end{figure}
Figure~\ref{fig:framework} decomposes each model's speedup into algorithmic and framework contributions.
For small blocks (GST-S), XLA kernel fusion yields $r \gg 1$, amplifying the algorithmic speedup.
GST-C2 and GST-C3 also benefit from $r > 1$ ($r{=}2.2$ and $2.8$); their short temporal loops ($n{=}8$--$12$) favor XLA's compiled scan over Serinv's per-iteration Python dispatch.
For compute-bound models (GST-M, GST-L, GST-T, SA1), both stacks dispatch to cuBLAS/cuSOLVER and $r \approx 0.9$--$1.0$, close to parity.
For distributed models, $r$ drops well below parity ($0.1$--$0.25$):
Serinv overlaps CPU$\leftrightarrow$GPU transfers with computation via direct CUDA stream control, whereas JAX's per-block JIT dispatch is synchronous.
This is a framework-level engineering limitation, not a fundamental cost of AD.

\subsection{Resource and Energy Efficiency}
\label{sec:eval:efficiency}

\researchquestion{FD's $2d{+}1$ evaluations are embarrassingly parallel. How many GPUs does FD need to match AD's speed, and at what energy cost?}

\smallskip

We measure per-gradient time for FD on increasing GPU counts.
FD scales by running its $2d{+}1$ evaluations in parallel: given $P$ GPUs, it runs $\lfloor P/P_{\min} \rfloor$ concurrent evaluations, each on $P_{\min}$ GPUs.
Since AD replaces these $2d{+}1$ evaluations with a single forward-backward pass, it is run on its minimum configuration only.
Energy is measured via Cray PM hardware counters at the node power-supply level (10\,Hz, capturing GPU, CPU, memory, and interconnect), with MPI barriers ensuring the measurement window includes idle time to reflect the real allocation cost.

\begin{figure*}[t]
    \centering
    \includegraphics[width=\textwidth]{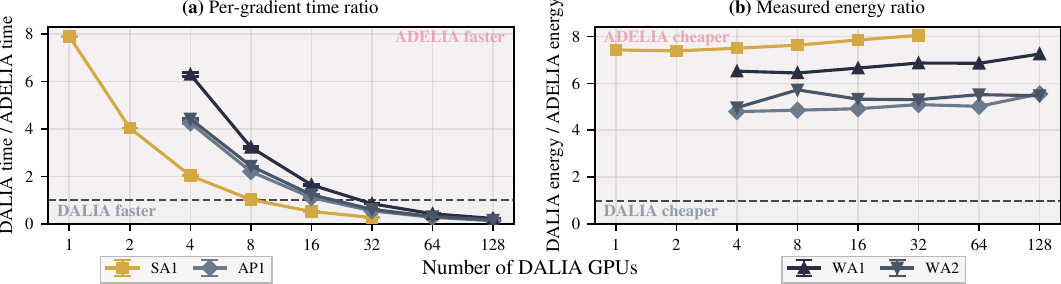}
    \vspace{-5mm}
    \caption{Resource and energy efficiency for the four production-scale models ($d{=}15$).
    (a)~Per-gradient speedup as FD scales to more GPUs; AD runs on $P_{\min}$.
    (b)~Energy ratio.
    Values above dashed lines favor AD.}
    \label{fig:efficiency}
    \vspace{-6mm}
\end{figure*}

\paragraph{Per-gradient time scaling}
Figure~\ref{fig:efficiency}(a) focuses on the four production-scale models, all with $d{=}15$ ($2d{+}1{=}31$ FD evaluations).
Under ideal parallelism, FD matches AD at $P^* \approx P_{\min} \cdot (2d{+}1)/c_{\mathrm{AD}}$ GPUs.
SA1 ($c_{\mathrm{AD}}{=}3.9$, single-GPU AD, $P_{\min}{=}1$) starts at $7.9\times$ with one FD GPU and crosses below~1 between 8 and 16~GPUs; its curve stops at 32~GPUs because $31 \leq 32/P_{\min}$, so all evaluations already run in a single round and additional GPUs sit idle.
The distributed models (AP1, WA1, WA2; $c_{\mathrm{AD}}{=}4.8$--$7.4$, 4-GPU AD) cross below~1 between 16 and 32~FD GPUs and are scaled to 128~GPUs.
For the smaller benchmarks (not shown in the figure), the pattern is similar: low-$d$ models ($d{=}4$) break even at 4--8~GPUs, while high-$d$ models ($d{=}9$--$15$) remain above break-even through 32~GPUs.

\paragraph{Energy efficiency}
Figure~\ref{fig:efficiency}(b) shows the opposite trend: unlike time ratios, which decrease as FD scales to more GPUs, energy ratios are flat or \emph{increasing}.
Under ideal parallelism, doubling GPUs halves FD's wall time but doubles the hardware, so total energy (  time $\times$ per-GPU power $\times$ GPUs) is conserved: parallelism trades time for GPUs, leaving total work unchanged.
In practice, sub-linear scaling and per-GPU base power draw makes the total energy grow with GPU count even as time decreases.

The best case for FD (WA1 at 128~GPUs, where FD is ${\sim}4.2\times$ faster), it still consumes $7.3\times$ more energy. Crucially, parallelism offers no remedy for gradient quality: The convergence gap shown in Section~\ref{sec:eval:gradval} persists regardless of scale. ADELIA is therefore the only path to both energy-efficient \emph{and} high-quality gradient computation.

\subsection{Performance Analysis}
\label{sec:eval:breakdown}

\researchquestion{Where do the time and memory costs concentrate in practice, and what are the practical deployment overheads?}

\begin{figure}[t]
    \centering
    \includegraphics[width=\linewidth]{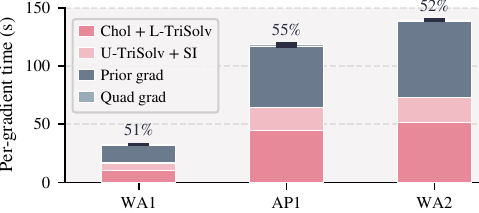}
    \vspace{-7mm}
    \caption{Per-stage time breakdown for distributed models on 4~GPUs. Percentages show the combined Python-loop fraction.}
    \label{fig:performance_analysis}
    \vspace{-6mm}
\end{figure}

\smallskip
\paragraph{Peak GPU memory}
ADELIA's peak GPU memory is dominated by the $n$ stored Schur complement carries ($b \times b$ each).
Total memory ranges from $<1$\,MiB (GST-S) to 63.3\,GiB (GST-L) for single-GPU models (Table~\ref{tab:structure_comparison}); GST-T uses 46.5\,GiB.
For distributed models, the CPU-staging strategy from Section~\ref{sec:method:multigpu} keeps per-GPU memory to a single block's workspace, with Schur carries stored in CPU memory.

\paragraph{Stage-level profiling}
Figure~\ref{fig:performance_analysis} breaks down the per-gradient time for the three distributed models into the four gradient stages from Section~\ref{sec:method:bta}.
The two Python-loop stages (Chol + L-TriSolv, U-TriSolv + SI) account for 51--55\% of total time; the prior log-determinant gradient (Phase~C) accounts for the remainder 44--47\%, scaling with block size.

\paragraph{Two-phase execution profile}
Phase~1 (Section~\ref{sec:method:multigpu}) dominates total time. Phase~2 overhead varies by model: negligible for WA1 (${\sim}7\%$, $b{=}3{,}741$) but ${\sim}24\%$ for AP1/WA2 ($b{>}12{,}000$), driven by the reduced system's allgather of $O(P \cdot b^2)$ elements (22--25\,s) and factorization (6--8\,s).

%% file: sections/related.tex
\section{Related Work}
\label{sec:related}

\paragraph{INLA Implementations}
R-INLA~\cite{rue2009approximate,gaedke2022parallelized} is the standard INLA implementation, using FD for gradients and PARDISO~\cite{schenk2004solving} for Cholesky and SI.
DALIA~\cite{dalia_sc} scales INLA to hundreds of GPUs by parallelizing across FD evaluations, precision matrix assembly, and distributed BTA solvers from Serinv~\cite{serinv}, but retains finite differences.
The gmrfs library~\cite{geraschenko2025gmrfs} uses JAX AD for INLA on GMRFs but differentiates through generic sparse Cholesky via Python loops without a custom backward pass, analogous to but less efficient than our AD-Loop baseline (Table~\ref{tab:structure_comparison}), which uses compiled \texttt{lax.scan}.
ADELIA replaces FD in DALIA with structure-exploiting AD while retaining the same solver infrastructure.

\paragraph{Sparse Solvers and AD for Structured Matrices}
General-purpose sparse direct solvers (PARDISO~\cite{schenk2004solving}, MUMPS~\cite{amestoy2001fully}, SuperLU\_DIST~\cite{li2003superlu}, cuSOLVER) use fill-reducing orderings that disrupt block-tridiagonal structure.
Serinv~\cite{serinv} operates on the BTA structure for $O(n \cdot b^3)$ factorization; Our carry-based reconstruction can be seen as a structure-aware form of gradient checkpointing~\cite{chen2016training}, exploiting block sparsity by storing Schur complement carries and reconstructing factors on the fly.
Durrande et al.~\cite{durrande2019banded} derived reverse-mode AD rules for banded Cholesky and Takahashi's equations on scalar banded matrices, but their approach does not handle block structure, arrowhead coupling, or GPU execution.
Algorithmically, the backward pass relates to the Rauch--Tung--Striebel smoother~\cite{rauch1965maximum}, but differentiable state-space models target dense states of tens to hundreds, whereas our spatial blocks reach $b{=}13{,}455$.

\paragraph{Probabilistic Programming and GPU Inference}
AD-based frameworks such as Stan~\cite{carpenter2017stan}, PyMC~\cite{abril2023pymc}, NumPyro~\cite{phan2019composable}, and TensorFlow Probability~\cite{dillon2017tensorflow} provide general Bayesian inference but do not exploit BTA sparsity; naive generic sparse AD destroys block structure.
TMB~\cite{kristensen2016tmb} combines tape-based AD (CppAD) with 
CHOLMOD's sparse Cholesky and the inverse subset algorithm to 
differentiate the Laplace-approximated marginal likelihood, supporting automatic sparsity detection for models with up to $10^6$ random effects. However, TMB uses general-purpose fill-reducing orderings that do not preserve BTA block structure, precluding the $O(n \cdot b^3)$ block-sequential algorithms central to our approach, and is limited to CPU execution.
Margossian et al.~\cite{margossian2020adjoint, margossian2023general} derive an adjoint-differentiated Laplace approximation that, like our approach, exploits the envelope theorem to avoid implicit differentiation through the mode; their 2023 generalization~\cite{margossian2023general} compares the inverse-subset and adjoint gradient strategies. The Stan implementation does not exploit structured sparsity and lacks GPU support, limiting scalability to the models supported by ADELIA.
None target INLA's combination of Laplace approximation, structured sparse factorization, and SI.

%% file: sections/conclusion.tex
\section{Conclusion}
\label{sec:conclusion}

We presented ADELIA, the first AD-enabled INLA framework that exploits the block sparsity of the underlying precision matrices to compute exact gradients on multiple GPUs, scaling to models with up to 1.9M latent variables. On four production-grade multivariate models, AD achieves $4.2$--$7.9\times$ per-gradient speedups on equal hardware, with $5$--$8\times$ energy savings even when FD matches wall-clock time by parallelizing over more nodes. Beyond performance, exact gradients are essential for the largest multivariate models, where FD stalls at gradient norms orders of magnitude above tolerance. Regardless of resource availability, ADELIA is the go-to option for fast, energy-efficient, accurate gradients for HPC-scale INLA applications. Enabling reliable convergence at these scales, at a cost independent of the number of hyperparameters, makes it possible to ask scientific questions that the FD computational overhead and approximation error previously prevented.

%% file: references.bib
@IEEEtranBSTCTL{BSTcontrol,
  CTLuse_url = {yes}
}

@book{gelman2013bayesian,
  author = {Gelman, Andrew and Carlin, John B. and Stern, Hal S. and Dunson, David B. and Vehtari, Aki and Rubin, Donald B.},
  title = {Bayesian Data Analysis},
  edition = {3rd},
  publisher = {Chapman and Hall/CRC},
  year = {2013},
  doi = {10.1201/b16018},
  url = {https://doi.org/10.1201/b16018},
}

@book{moraga2019geospatial,
  title={Geospatial Health Data: Modeling and visualization with R-INLA and Shiny},
  author={Moraga, Paula},
  year={2019},
  edition = {1st},
  publisher={Chapman and Hall/CRC},
  doi={10.1201/9780429341823},
  url = {https://doi.org/10.1201/9780429341823},
}

@book{wikle2019spatio,
  title={Spatio-temporal statistics with R},
  author={Wikle, Christopher K and Zammit-Mangion, Andrew and Cressie, Noel},
  year={2019},
  edition = {1st},
  publisher={Chapman and Hall/CRC},
  doi={10.1201/9781351769723},
  url = {https://doi.org/10.1201/9781351769723},
}

@book{blangiardo2015spatial,
  title={Spatial and spatio-temporal Bayesian models with R-INLA},
  author={Blangiardo, Marta and Cameletti, Michela},
  year={2015},
  publisher={John Wiley \& Sons, Ltd},
  doi={10.1002/9781118950203},
  isbn = {9781118950203},
  url = {https://doi.org/10.1002/9781118950203},
}

@book{brooks2011handbook,
  author = {Brooks, Steve and Gelman, Andrew and Jones, Galin and Meng, Xiao-Li},
  title = {Handbook of Markov Chain Monte Carlo},
  publisher = {Chapman and Hall/CRC},
  year = {2011},
  edition = {1st},
  doi = {10.1201/b10905},
  url = {https://doi.org/10.1201/b10905},
}

@article{rue2009approximate,
  author = {Rue, H{\aa}vard and Martino, Sara and Chopin, Nicolas},
  title = {Approximate Bayesian Inference for Latent Gaussian models by using Integrated Nested Laplace Approximations},
  journal = {Journal of the Royal Statistical Society Series B: Statistical Methodology},
  volume = {71},
  number = {2},
  pages = {319-392},
  year = {2009},
  month = {04},
  issn = {1369-7412},
  doi = {10.1111/j.1467-9868.2008.00700.x},
  url = {https://doi.org/10.1111/j.1467-9868.2008.00700.x},
  eprint = {https://academic.oup.com/jrsssb/article-pdf/71/2/319/49686253/jrsssb_71_2_319.pdf}
}

@article{lindgren2011explicit,
  author = {Lindgren, Finn and Rue, H{\aa}vard and Lindstr{\"o}m, Johan},
  title = {An explicit link between {Gaussian} fields and {Gaussian} {Markov} random fields: the stochastic partial differential equation approach},
  journal = {Journal of the Royal Statistical Society Series B: Statistical Methodology},
  volume = {73},
  number = {4},
  pages = {423-498},
  year = {2011},
  month = {08},
  doi = {10.1111/j.1467-9868.2011.00777.x},
  url = {https://doi.org/10.1111/j.1467-9868.2011.00777.x},
}

@book{rue2005gaussian,
  author = {Rue, H{\aa}vard and Held, Leonhard},
  title = {Gaussian {Markov} Random Fields: Theory and Applications},
  publisher = {Chapman and Hall/CRC},
  year = {2005},
  doi = {10.1201/9780203492024},
  url = {https://doi.org/10.1201/9780203492024},
}

@book{nocedal2006numerical,
  author = {Nocedal, Jorge and Wright, Stephen J.},
  title = {Numerical Optimization},
  edition = {2nd},
  publisher = {Springer},
  year = {2006},
  doi = {10.1007/978-0-387-40065-5},
  url = {https://doi.org/10.1007/978-0-387-40065-5},
}

@misc{lindgren2022diffusion,
	author = {Finn Lindgren and Haakon Bakka and David Bolin and Elias Krainski and H{\aa}vard Rue},
	title =	 {A diffusion-based spatio-temporal extension of Gaussian Mat\'ern fields},
    year={2023},
    eprint={2006.04917},
	archivePrefix={arXiv},
	doi={10.48550/arXiv.2006.04917},
  url = {https://doi.org/10.48550/arXiv.2006.04917},
}

@article{gaedkeIntegrated2024,
	author = {Gaedke-Merzh\"{a}user, Lisa and Krainski, Elias and Janalik, Radim and Rue, H\r{a}vard and Schenk, Olaf},
	title = {Integrated Nested Laplace Approximations for Large-Scale Spatiotemporal Bayesian Modeling},
	journal = {SIAM Journal on Scientific Computing},
	volume = {46},
	number = {4},
	pages = {B448-B473},
	year = {2024},
	doi = {10.1137/23M1561531},
  url = {https://doi.org/10.1137/23M1561531},
}

@article{gaedke2022parallelized,
	title={{Parallelized integrated nested {L}}aplace approximations for fast {Bayesian} inference},
	author={Gaedke-Merzh{\"a}user, Lisa and van Niekerk, Janet and Schenk, Olaf and Rue, H{\aa}vard},
	journal={Statistics and Computing},
	volume={33},
	pages={25},
	year={2023},
	doi={10.1007/s11222-022-10192-1},
  url = {https://doi.org/10.1007/s11222-022-10192-1},
}

@book{griewank2008evaluating,
  author = {Griewank, Andreas and Walther, Andrea},
  title = {Evaluating Derivatives},
  edition = {Second},
  publisher = {Society for Industrial and Applied Mathematics},
  year = {2008},
  address = {},
  doi = {10.1137/1.9780898717761},
  url = {https://doi.org/10.1137/1.9780898717761},
}

@article{baydin2018automatic,
  author = {Atilim Gunes Baydin and Barak A. Pearlmutter and Alexey Andreyevich Radul and Jeffrey Mark Siskind},
  title = {Automatic differentiation in machine learning: a survey},
  year={2018},
  eprint={1502.05767},
  archivePrefix={arXiv},
  primaryClass={cs.SC},
  year = {2018},
  doi = {10.48550/arXiv.1502.05767},
  url = {https://doi.org/10.48550/arXiv.1502.05767}
}

@misc{jax2018github,
  author = {James Bradbury and Roy Frostig and Peter Hawkins and Matthew James Johnson and Yash Katariya and Chris Leary and Dougal Maclaurin and George Necula and Adam Paszke and Jake Vander{P}las and Skye Wanderman-{M}ilne and Qiao Zhang},
  title = {{JAX}: composable transformations of {Python}+{NumPy} programs},
  url = {http://github.com/jax-ml/jax},
  version = {0.3.13},
  year = {2018}
}

@misc{chen2016training,
  author = {Tianqi Chen and Bing Xu and Chiyuan Zhang and Carlos Guestrin},
  title = {Training deep nets with sublinear memory cost},
  year = {2016},
  eprint = {1604.06174},
  archivePrefix = {arXiv},
  primaryClass = {cs.LG},
  doi = {10.48550/arXiv.1604.06174},
  url = {https://doi.org/10.48550/arXiv.1604.06174}
}

@misc{petersen2012matrix,
  author = {Petersen, Kaare Brandt and Pedersen, Michael Syskind},
  title = {The Matrix Cookbook},
  year = {2012},
  publisher = {Technical University of Denmark},
  note = {Version 20121115},
  url = {http://www2.compute.dtu.dk/pubdb/pubs/3274-full.html}
}

@misc{murray2016differentiation,
  author = {Murray, Iain},
  title = {Differentiation of the {Cholesky} decomposition},
  year = {2016},
  eprint = {1602.07527},
  archivePrefix = {arXiv},
  primaryClass = {stat.CO},
  doi = {10.48550/arXiv.1602.07527},
  url = {https://doi.org/10.48550/arXiv.1602.07527}
}

@article{abril2023pymc,
  author = {Abril-Pla, Oriol and Andreani, Virgile and Carroll, Colin and Dong, Larry and Fonnesbeck, Christopher J. and Kochurov, Maxim and Kumar, Ravin and Lao, Junpeng and Luhmann, Christian C. and Martin, Osvaldo A. and Osthege, Michael and Vieira, Ricardo and Wiecki, Thomas and Zinkov, Robert},
  title = {{PyMC}: a modern, and comprehensive probabilistic programming framework in {Python}},
  journal = {PeerJ Computer Science},
  volume = {9},
  pages = {e1516},
  year = {2023},
  doi = {10.7717/peerj-cs.1516},
  url = {https://doi.org/10.7717/peerj-cs.1516},
}

@article{carpenter2017stan,
  author = {Carpenter, Bob and Gelman, Andrew and Hoffman, Matthew D. and Lee, Daniel and Goodrich, Ben and Betancourt, Michael and Brubaker, Marcus and Guo, Jiqiang and Li, Peter and Riddell, Allen},
  title = {Stan: A probabilistic programming language},
  journal = {Journal of Statistical Software},
  volume = {76},
  number = {1},
  pages = {1-32},
  year = {2017},
  doi = {10.18637/jss.v076.i01},
  url = {https://doi.org/10.18637/jss.v076.i01},
}

@inproceedings{phan2019composable,
  author = {Du Phan and Neeraj Pradhan and Martin Jankowiak},
  title = {Composable effects for flexible and accelerated probabilistic programming in {NumPyro}},
  booktitle = {NeurIPS Workshop on Program Transformations},
  year = {2019},
  doi = {10.48550/arXiv.1912.11554},
  url = {https://doi.org/10.48550/arXiv.1912.11554}
}

@misc{dillon2017tensorflow,
  author = {Joshua V. Dillon and Ian Langmore and Dustin Tran and Eugene Brevdo and Srinivas Vasudevan and Dave Moore and Brian Patton and Alex Alemi and Matt Hoffman and Rif A. Saurous},
  title = {{TensorFlow} Distributions},
  year = {2017},
  eprint = {1711.10604},
  archivePrefix = {arXiv},
  primaryClass = {cs.LG},
  doi = {10.48550/arXiv.1711.10604},
  url = {https://doi.org/10.48550/arXiv.1711.10604}
}

@inproceedings{hoefler2015scientific,
  author = {Hoefler, Torsten and Belli, Roberto},
  title = {Scientific Benchmarking of Parallel Computing Systems: Twelve Ways to Tell the Masses When Reporting Performance Results},
  publisher = {Association for Computing Machinery},
  booktitle = {Proceedings of the International Conference for High Performance Computing, Networking, Storage and Analysis (SC)},
  articleno = {73},
  numpages = {12},
  year = {2015},
  doi = {10.1145/2807591.2807644},
  url = {https://doi.org/10.1145/2807591.2807644},
  location = {Austin, Texas},
  series = {SC '15}
}

@article{schenk2004solving,
  author = {Schenk, Olaf and G{\"a}rtner, Klaus},
  title = {Solving unsymmetric sparse systems of linear equations with {PARDISO}},
  journal = {Future Generation Computer Systems},
  volume = {20},
  number = {3},
  pages = {475-487},
  year = {2004},
  note = {Selected numerical algorithms},
  doi = {10.1016/j.future.2003.07.011},
  url = {https://doi.org/10.1016/j.future.2003.07.011},
}

@article{amestoy2001fully,
  author = {Amestoy, Patrick R. and Duff, Iain S. and L'Excellent, Jean-Yves and Koster, Jacko},
  title = {A fully asynchronous multifrontal solver using distributed dynamic scheduling},
  journal = {SIAM Journal on Matrix Analysis and Applications},
  volume = {23},
  number = {1},
  pages = {15-41},
  year = {2001},
  doi = {10.1137/S0895479899358194},
  url = {https://doi.org/10.1137/S0895479899358194},
}

@article{li2003superlu,
  author = {Li, Xiaoye S. and Demmel, James W.},
  title = {{SuperLU\_DIST}: A scalable distributed-memory sparse direct solver for unsymmetric linear systems},
  publisher = {Association for Computing Machinery},
address = {New York, NY, USA},
  journal = {ACM Trans. Math. Softw.},
  volume = {29},
  number = {2},
  pages = {110-140},
  year = {2003},
  doi = {10.1145/779359.779361},
  url = {https://doi.org/10.1145/779359.779361},
}

@article{rauch1965maximum,
  author = {RAUCH, H. E. and TUNG, F. and STRIEBEL, C. T.},
  title = {Maximum likelihood estimates of linear dynamic systems},
  journal = {AIAA Journal},
  volume = {3},
  number = {8},
  pages = {1445-1450},
  year = {1965},
  doi = {10.2514/3.3166},
  url = {https://doi.org/10.2514/3.3166},
}

@misc{dalia_sc,
  title={Accelerated Spatio-Temporal Bayesian Modeling for Multivariate Gaussian Processes},
  author={Lisa Gaedke-Merzh{\"a}user and Vincent Maillou and Fernando Rodriguez Avellaneda and Olaf Schenk and Mathieu Luisier and Paula Moraga and Alexandros Nikolaos Ziogas and H{\aa}vard Rue},
  year={2025},
  eprint={2507.06938},
  archivePrefix={arXiv},
  primaryClass={stat.CO},
  url={https://doi.org/10.48550/arXiv.2507.06938},
  doi={10.48550/arXiv.2507.06938}
}

@misc{serinv,
  title={Serinv: A Scalable Library for the Selected Inversion of Block-Tridiagonal with Arrowhead Matrices},
  author={Vincent Maillou and Lisa Gaedke-Merzh{\"a}user and Alexandros Nikolaos Ziogas and Olaf Schenk and Mathieu Luisier},
  year={2025},
  eprint={2503.17528},
  archivePrefix={arXiv},
  primaryClass={cs.DC},
  url={https://doi.org/10.48550/arXiv.2503.17528},
  doi={10.48550/arXiv.2503.17528}
}

@article{mpi4jax,
  author = {H{\"a}fner, Dion and Vicentini, Filippo},
  title = {mpi4jax: Zero-copy {MPI} communication of {JAX} arrays},
  journal = {Journal of Open Source Software},
  volume = {6},
  number = {65},
  pages = {3419},
  year = {2021},
  doi = {10.21105/joss.03419},
  url = {https://doi.org/10.21105/joss.03419},
}

@misc{durrande2019banded,
  author = {Nicolas Durrande and Vincent Adam and Lucas Bordeaux and Stefanos Eleftheriadis and James Hensman},
  title = {Banded Matrix Operators for {Gaussian} {Markov} Models in the Automatic Differentiation Era},
  year = {2019},
  eprint = {1902.10078},
  archivePrefix = {arXiv},
  primaryClass = {stat.ML},
  doi = {10.48550/arXiv.1902.10078},
  url = {https://doi.org/10.48550/arXiv.1902.10078},
}

@article{kristensen2016tmb,
  author = {Kristensen, Kasper and Nielsen, Anders and Berg, Casper W. and Skaug, Hans and Bell, Bradley M.},
  title = {{TMB}: Automatic Differentiation and {Laplace} Approximation},
  journal = {Journal of Statistical Software},
  volume = {70},
  number = {5},
  pages = {1–21},
  year = {2016},
  doi = {10.18637/jss.v070.i05},
  url = {https://doi.org/10.18637/jss.v070.i05},
}

@inproceedings{margossian2020adjoint,
  author = {Charles C. Margossian and Aki Vehtari and Daniel Simpson and Raj Agrawal},
  title = {Hamiltonian {Monte} {Carlo} using an adjoint-differentiated {Laplace} approximation: {Bayesian} inference for latent {Gaussian} models and beyond},
  eprint={2004.12550},
  archivePrefix={arXiv},
  primaryClass={stat.CO},
  year = {2020},
  url = {https://doi.org/10.48550/arXiv.2004.12550}
}

@misc{margossian2023general,
  author = {Margossian, Charles C.},
  title = {General adjoint-differentiated {Laplace} approximation},
  year = {2023},
  eprint = {2306.14976},
  archivePrefix = {arXiv},
  primaryClass = {stat.CO},
  doi = {10.48550/arXiv.2306.14976},
  url = {https://doi.org/10.48550/arXiv.2306.14976},
}

@article{milgrom2002envelope,
  author = {Milgrom, Paul and Segal, Ilya},
  title = {Envelope Theorems for Arbitrary Choice Sets},
  journal = {Econometrica},
  volume = {70},
  number = {2},
  pages = {583-601},
  year = {2002},
  doi = {10.1111/1468-0262.00296},
  url = {https://doi.org/10.1111/1468-0262.00296},
}

@misc{geraschenko2025gmrfs,
  title = {gmrfs: {INLA} for {Gaussian} {Markov} Random Fields in {JAX}},
  author = {Geraschenko, Anton},
  year = {2024},
  url = {https://github.com/geraschenko/gmrfs}
}

@inproceedings{10.1145/3757348.3757365,
  author = {Martinasso, Maxime and Klein, Mark and Schulthess, Thomas},
  title = {Alps, a versatile research infrastructure},
  year = {2025},
  isbn = {9798400713279},
  publisher = {Association for Computing Machinery},
  address = {New York, NY, USA},
  url = {https://doi.org/10.1145/3757348.3757365},
  doi = {10.1145/3757348.3757365},
  booktitle = {Proceedings of the Cray User Group},
  pages = {156-165},
  numpages = {10},
  series = {CUG '25}
}

@article{schmidt2003bayesian,
  title={A Bayesian coregionalization approach for multivariate pollutant data},
  author={Schmidt, Alexandra M and Gelfand, Alan E},
  journal={Journal of Geophysical Research: Atmospheres},
  volume={108},
  number={D24},
  year={2003},
  publisher={Wiley Online Library},
  url = {https://doi.org/10.1029/2002JD002905}
}
